\documentclass{ws-ijqi}
\newcommand{\ket}[1]{\mbox{$|#1\rangle$}}

\newcommand{\vect}[1]{\underline #1}
\begin{document}

\markboth{F. Bal\'azs \& S. Imre} {Quantum Computation Based PDF
Estimation}

\catchline{}{}{}{}{}

\title{Quantum Computation Based Probability Density Function Estimation}

\author{Ferenc Bal\'azs, S\'andor Imre}
\address{Mobile Communications \& Computing Laboratory\\
Department of Telecommunications\\  Budapest University of
Technology and Economics\\ 1117 Budapest, Magyar Tud\'osok krt.
2, HUNGARY\\
\scriptsize\tt{email: \{balazsf,imre\}@hit.bme.hu}}



\maketitle


\begin{abstract}
Signal processing techniques will lean on blind methods in the
near future, where no redundant, resource allocating information
will be transmitted through the channel. To achieve a proper
decision, however, it is essential to know at least the
probability density function (pdf), which to estimate is
classically a time consumpting and/or less accurate hard task,
that may make decisions to fail. This paper describes the design
of a quantum assisted pdf estimation method also by an example,
which promises to achieve the exact pdf by proper setting of
parameters in a very fast way.
\end{abstract}



\section{Introduction}
The latest prognoses forecast a transistor size shrinkage to its 1
nm limit up to the year 2010, where the impacts of the quantum
effects rise remarkably\cite{cle04}. Several new methods were
published to replace the silicon based technology, one of them
exploits the power of quantum computation. Unfortunately, it does
not exists quantum computer solution with proper size, so far. In
addition several research papers showed the constraints of the
application range of quantum computers, that would not relegate
classical computation which shows the demand on quantum assisted
computation, which could amplify computational power enormously in
the next future. This research work focuses on the exploitation of
the strengths of quantum computation in classically difficult
problems. \par In every communication chain a decision has to be
made at least in the receiver side, often without any
\textit{a-priori} known information, e.g. \textit{a-posteriori}
probability or the probability density function (pdf) of the
source.
\par The estimation of the probability density function of
observed signals shows a significant interest in many signal
processing methods, such as pattern recognition, independent
component analysis or detection. It exists a lot of well defined
estimation techniques based on histogram or kernel
method\cite{dev85}, which requires a large number of samples $n$
to estimate the pdf almost sure as described for the $L_1$ case by
\begin{equation}
\lim_{n\rightarrow\infty}\|f-f_n\|=\lim_{n\rightarrow\infty}\int\left|f(x)-f_n(x)\right|\mathrm{d}x\rightarrow
0. \label{eq:1}
\end{equation}
A more tight lower bound for kernel estimator can be found, e.g.
in\cite{Dev89}. In case of decision problems, however, the
knowledge of the whole pdf is often not needed, rather the pdf at
a single point is of interest.
\par The rest of the paper is organized as follows: the overall
system model is introduced in Section~\ref{sec:system}. In
Section~\ref{sec:method} the probability density function
estimation is showed, and Section~\ref{sec:conc} concludes the
paper.

\section{System Model}\label{sec:system}
\par By help of \textit{a-priori} knowledge of the observed system\footnote{This is not an \textit{a-priori} information about the sequence addressed above.}, a quantum register
\begin{equation}
\ket{\varphi}=\sum_{x=0}^{N-1}\varphi_x\ket{x}; ~~~\varphi_x\in
\mathbb{C}, \label{eq:2}\end{equation} as a database is used to
store all the raw quantized parameters, e.g. delay, heat,
velocity, etc. values\cite{Imr01c}. The database (\ref{eq:2}) have
to be set up in the hardware only once for the whole estimation
process. To handle the large amount of data a virtual database
$y=g(s,\vect{x})$, 
should be introduced\cite{Imr04}, where $s$ describes the behavior
of the system and $\vect{x}$ denotes the index of the qregister
$\ket{\varphi}$, respectively. The function $y_i=g(s,x_i)$ points
to an record in the virtual database\cite{Imr01c}.

\subsection{Properties of the Virtual Database Generating Function}\label{ssec:prop} The
function $g(s,\vect{x})$ is not obligingly mutual unambiguous
consequently, i.e. it is not reversible, except for several special
cases, when the virtual database contains
$\widehat{r}=g(s,\vect{x})$ only once. In this case the parameter
settings of the system are easy to determine. Henceforth the fact
should be kept in mind that $g(s,\vect{x})$ is in almost every case
a so called one way function which is easy to evaluate in one
direction, but to estimate the inverse is rather hard. \par The
function $g(.)$ generates all the possible disturbances additional
to the considered input value. This is of course a large amount of
information, $2N=2^{n+1}$, where $n$ is the length of the qregister
$\ket{\varphi}$. For an example let us assume a 15-qbit qregister.
The function $g(\cdot)$ in (\ref{eq:2}) generates $2^{15}=32.768$
output values. Taking into account the large number of possible
points in the set surface the optimal classification in a classical
way becomes difficult. At the first glace this problem looks more
difficult to solve, however, using the Deutsch-Jozsa\cite{Deu92}
quantum parallelization algorithm, an arbitrary unitary operation
can be executed on all the prepared states contemporaneously.
\subsection{The Estimator} 
\begin{figure}[tb]
\begin{center}
\includegraphics[width=50mm, height=25mm]{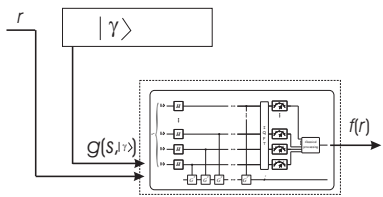}
\end{center}
\caption{Pdf estimator}\label{fig:ijqi1}
\end{figure}

Roughly speaking the task is to find the
entry (entries) in the virtual databases which is (are) equal to the
observed data $r$. To accomplish the database search the Grover
database search algorithm should be invoked\cite{Gro96}, where we
feed the received signal $r(t)$ and $g(s,\vect{x})$ to the oracle
$(\mathcal{O})$. Because of the fact of tight bound, in real
application less iterations would be also appropriate\cite{Imr03}.
Employing the Grover database search algorithm we are able to find
the entries in the virtual databases, however, it is not needed to
perform a complete search because the search result --the exact
index (indices) of the searched item(s)-- is (are) not interesting
but the number how often a given configuration is involved in
$g(s,\vect{x})$ or not. For that purpose a new function
\begin{equation}
f(r|s)=\frac{\sharp\left(x: r=g(s,x)\right)}{\sharp (x)},
\label{eq:4}
\end{equation}
is defined\cite{Imr04}, which counts the number of similar entries
in the virtual database, which corresponds to the conditional
probability density function $r$ to be in the set $s$. For that
reason it is worth stepping forward to quantum
counting\cite{Bra98_2} based on Grover iteration
(Fig.~\ref{fig:ijqi1}).
\section{Estimation Method}\label{sec:method}In this section we deal with the
estimation of the probability density function at a single input
point and with the estimation of the whole pdf by quantum assisted
way.
\subsection{Decision in One Point}\label{ssec:3.1}
\begin{figure}[tb]
\begin{center}
\includegraphics[width=55mm, height=35mm]{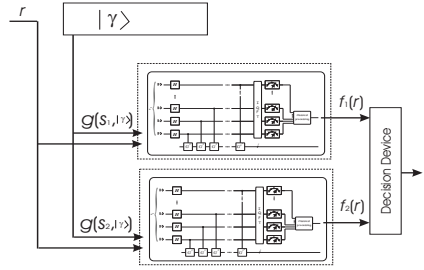}
\end{center}
\caption{Quantum Assisted Detector}\label{fig:ijqi2}
\end{figure}

\begin{figure}[b]
\begin{center}
\includegraphics[width=38mm, height=26mm]{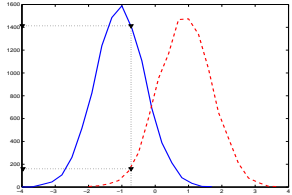}
\end{center}
\caption{Calculated appearance of possible input signals in
Gaussian distributed noisy environment}\label{fig:pdf2}
\end{figure}

Based on the proposed pdf estimation method a decision is possible
at a single input point. As an example a decision has to be made
in a telecommunication receiver. A received signal $r$ is given,
which is the Bernoulli-distributed source signal $a\in
\mathbb{R}^{[0,1]}$ disturbed by the communication channel. For
the sake of simplicity only an Additive White Gaussian Noise
(AWGN) channel is assumed, $r(t)=a(t)+n(t)$, where $n$ is the
Gaussian distributed noise sample. In case of no known
\textit{a-priori} probability a decision in Maximum Likelihood
(ML) sense seems to be appropriate, which maximize the likelihood
function based on the (estimated) pdf. The value $r=-0.8$ was
received in the detector, the task is to decide whether $a=-1$
with noise vector $n=0.2$ or $a=1$ with $n=-1.8$ was detected as
shown in Fig~\ref{fig:pdf2}. Employing the pdf estimator and the
decision device (a simple comparator) as shown in
Fig.~\ref{fig:ijqi2} the number of the overall accuracies of the
given $r$ in the system is calculated, which is slightly more than
1400 for the $a=-1$ belonging to the solid lined pdf, and less
than 200 for the other one. In this way the likelihood functions
are well calculated and a fast decision is possible. You may find
another example in\cite{Imr04}.

\subsection{The Whole Pdf Estimation}
In many cases, however, the knowledge of the whole pdf is
required, e.g. for independence testing of input sequences by KL
divergence
\begin{equation*}
KL(X\|Y)=P\left(X\log\left(\frac{X}{Y}\right)\right),
\end{equation*}
where $X=\prod_iY_i$ have to be fulfilled for the independence. To
get know the pdf in quantum assisted way a estimation method
introduced in (\ref{ssec:3.1}) should be invoked for a large
number of input values ($r$) to get a good approximation
\begin{equation}
\lim_{j\rightarrow
J}\left\|f(x_j|s)_{|x=r}-f(x)\right\|\rightarrow 0,
\end{equation}
in $L_1$ sense. Should be $J$ chosen to $\infty$ the method may
converge to continuous pdf estimation, which is impossible because
of time consumption and quantized behavior of the element in the
database (\ref{eq:2}). The problem is quite similar to the problem
of the number of bins in classical histogram method\cite{Dev01}.

\section{Concluding Remarks}\label{sec:conc}
This paper provides a brief introduction to quantum assisted
probability density function estimation. The inaccuracy of the pdf
estimation may result in wrong detection, however, the quantum
assisted estimator is able to achieve the exact value of the pdf
at a single point (received signal) that could make a decision
more accurate. The proposed qregister $\ket{\varphi}$ have to be
set up only once before the estimation. The virtual databases are
generated once and directly leaded to the Grover block in the
quantum counting circuit, which reduce the computational
complexity, substantially. The paper should be regarded as a
starting point to further analyze the properties of the quantum
assisted pdf estimator.


\section*{Acknowledgements}

The research project was partly supported by OTKA, id. Nr.:
F042590 and IST-2001-37944 FP5 Nexway project.








\end{document}